\documentstyle[11pt]{article}

\def\kon#1#2{\vbox{\halign{##&&##\cr\lower4pt
\hbox{$\scriptscriptstyle\vert$}\hrulefill &\hrulefill\lower4pt
\hbox{$\scriptscriptstyle\vert$}\cr $#1$&$#2$\cr}}}

\def\fii{\varphi}
\def\al{\alpha}

\def\eh{{\scriptstyle{1\over 2}}}
\def\d{\partial}
\def\=d{\,{\buildrel\rm def\over =}\,}

\def\sqr#1#2{{\vcenter{\vbox{\hrule height.#2pt\hbox{\vrule width.
#2pt height#1pt \kern#1pt \vrule width.#2pt}\hrule height.#2pt}}}}

\def\te{\vartheta}
\def\B{\Bigl}
\def\cost{{\rm const.}}

\begin{document}

\title{Against geometry: Nonstandard general relativity }
\author{G\"unter Scharf \footnote{e-mail: scharf@physik.uzh.ch}
\\ Institut f\"ur Theoretische Physik, 
\\ Universit\"at Z\"urich, 
\\ Winterthurerstr. 190 , CH-8057 Z\"urich, Switzerland}

\date{}

\maketitle\vskip 3cm

\begin{abstract}  
We show that the Schwarzschild solution can be embedded in a class of nonstandard solutions of the vacuum Einstein's equations with arbitrary rotation curves. These nonstandard solutions have to be taken as physical, if dark matter as needed in the standard theory cannot be found. As a
consequence general relativity is considered as a classical field theory in Minkowski space and not as a geometric theory in the sense of Einstein. Assuming an asymptotically flat rotation curve and introducing a material disk into this model we find a matter density in accordance with the Tully-Fisher relation.

\end{abstract}

\newpage

\section{Introduction}

General relativity is a classical gauge theory. This implies that the fundamental fields, as the metric $g_{\mu\nu}$, are not directly observable. Therefore, in investigating gravitational effects it is important to identify the observable quantities which are actually measurable. Every observable is defined by a measuring process, the coordinate system included. A change of the coordinates, although mathematically possible, is dangerous for physical reasons: one may lose the contact with the measurable observables. Therefore, we choose physically defined coordinates once and for all and do not change them. On the scale of galaxies one most important observable is the circular velocity $V(r)$ of stars or gas which can be measured by the Doppler shift of spectral lines ($r$ is the radius of the circular orbit). This observable plays an important role in the following: we will use it to fix the gauge. 

In standard general relativity one is tempted to interpret the metric $g_{\mu\nu}$ geometrically, for example by using it to measure the circumference of a circle in space. We reject this because, similarly as in electrodynamics, the $g_{\mu\nu}$ are the gravitational {\it potentials} and as such they are not observable. A nice way to see this is to consider electrodynamics and general relativity side by side. The electromagnetic fields $F^{\mu\nu}$ are defined by their effect on the motion of charged test bodies according to the equation of motion
$${d^2x^\mu\over d\tau^2}={e\over m}F^{\mu\nu}{dx_\nu\over d\tau},\eqno(1.1)$$
here the Lorentz force appears on the r.h.s. The corresponding equation of motion for test bodies in a gravitational field is the geodesic equation
$${d^2x^\mu\over d\tau^2}+\Gamma^\mu_{\al\beta}{dx^\al\over d\tau}{d x^\beta\over d\tau}=0.\eqno(1.2)$$
Consequently the Christoffel symbols are the gravitational field strengths. The field equations for the $F^{\mu\nu}$ are the inhomogeneous Maxwell's equations
$$\d_\nu F^{\mu\nu}=-\mu_0 j^\mu.\eqno(1.3)$$
The field equations for the $\Gamma^\mu_{\alpha\beta}$ are Einstein's equations
$$R_{\mu\nu}=\kappa(T_{\mu\nu}-\eh g_{\mu\nu}T_\alpha^\alpha).\eqno(1.4)$$
These are first order partial differential equations as (1.3), because the Ricci tensor is given by
$$R_{\mu\nu}=\d_\al \Gamma^\al_{\mu\nu}-\d_\nu \Gamma^\al_{\mu\al}+\Gamma^\al_{\al\beta}\Gamma^\beta_{\mu\nu}
-\Gamma^\al_{\nu\beta}\Gamma^\beta_{\al\mu}.\eqno(1.5)$$
However (1.3) are only four equations for the 6 components of $F^{\mu\nu}$. The gap is filled by introducing the vector
potential
$$F^{\mu\nu}=\d^\mu A^\nu-\d^\nu A^\mu,\eqno(1.6)$$
which is a consequence of the homogeneous Maxwell's equations. Similarly (1.4) are only 10 equations for the 40 components of $\Gamma$. The gap is filled by introducing the metric according to
$$\Gamma^\al_{\beta\gamma}={1\over 2}g^{\al\mu}(\d_\gamma g_{\beta\mu}+\d_\beta g_{\mu\gamma}-\d_\mu g_{\beta\gamma}).
\eqno(1.7)$$
Since the electromagnetic potentials are not observable quantities, the same must be true for the metric tensor which, therefore, has no direct physical interpretation in general. Consequently, {\it we do not interprete $g_{\mu\nu}$ geometrically; it is a parametrization of the gravitational field and nothing else.} Here we are following Poincar\'e ([1], p.50) and consider geometry as a convention. Standard general relativity is based on the fusion of geometry and gravitation, and one has considered this fusion as the most beautiful achievement of the general theory of relativity (W. Pauli, The Theory of Relativity, Dover, p.148). In the nonstandard theory this fusion is suspended.

In the theoretical analysis one should try to relate the observables to the metric $g_{\mu\nu}(x)$. The best would be if the metric under certain assumptions can be uniquely expressed by the observables. Then the gauge ambiguity has been removed, the gauge is fixed by a physical requirement. Another possibility is to choose the gauge on unphysical grounds, for example by some geometric convention or/and to simplify the solution of the differential equations. This standard approach is dangerous because one might miss some important physics. Our program of fixing the gauge by observables
is a sort of inverse procedure compared with standard general relativity where one first calculates a metric by solving Einstein's equations in some special gauge and then determines the observables. Clearly, in standard general relativity one cannot be sure that one finds all physically relevant solutions. Indeed we are going to show that Einstein's equations have {\it vacuum} solutions with an asymptotically flat rotation curve $V(r)$. These nonstandard solutions can be used to describe the dark halo of galaxies without introducing hypothetical dark matter. After the recent Xenon-experiment has again not found any signal of dark matter particles [2] one should seriously investigate nonstandard general relativity. 

The paper is organized as follows. As preliminaries we first consider the rotation curve in a general spherically symmetric gravitational field. Although this may be known we do not know a good reference. In Sect.3 we solve the vacuum Einstein's equations in our general spherically symmetric setting and express the metric tensor by the circular velocity $V(r)$. We find a class of nonstandard solutions which contains the Schwarzschild solution as a special case. All these solutions describe different physics if the circular velocities are different. From these vacuum solutions we construct in Sect.4 solutions with a disk of ordinary matter by means of the well-known displace, cut, and reflect method. This gives a simple model of a spiral galaxy. Assuming a circular velocity which is constant $=V_{\rm flat}$ for large $r$, we find a matter density proportional to $V_{\rm flat}^4$. This is in accordance with the Tully-Fisher relation [4] [5]. We close with some concluding remarks about standard and nonstandard general relativity. In particular we discuss the connection with MOND.

\section{The circular velocity in general relativity}

Standard geometric general relativity describes the solar system very well. But on the scale of galaxies which is $10^8$ times bigger one observes circular velocities of stars and gas much too big. These velocities are found from redshift measurements according to the formula for the special relativistic Doppler effect
$${\nu_{\rm obs}\over\nu}=(1+V_r)^{-1}(1-\vec V^2)^{-1/2}.\eqno(2.1)$$
Here $\nu$ is the frequency of a spectral line as known from atomic physics and $\nu_{\rm obs}$ actually measured with the telescope; $\vec V$ is the velocity of the light source and $V_r$ the component in the direction from observer to light source. We emphasize that $\vec V$ is a 3-vector but not a 4-tensor. Therefore, it changes under coordinate transformations, so it must be calculated in the rest system of the observer.

We consider a star moving in a static spherically symmetric gravitational field with the metric
$$ds^2=g_{\mu\nu}dx^\mu dx^\nu=e^a dt^2-e^bdr^2-r^2e^c(d\te^2+\sin^2\te d\phi^2),\eqno(2.2)$$
where $a(r), b(r), c(r)$ are functions of $r$ only. The coordinates $x^\mu=(t,r,\te,\phi)$ are measured in the laboratory system which is attached to the astronomers telescope. In standard geometric general relativity one says that $r$ is the ``circumference radius'' and so puts $c=0$. In our non-geometric interpretation of the $g_{\mu\nu}$ in (2.2) as a gravitational potentials there is no reason to do so. On the other hand a coordinate transformation $\bar r=\bar r(r)$ which removes $\exp c(r)$ is not allowed because the new radial coordinate would be unphysical. We shall return to this important point  at the end of the next section. 

We now calculate the gravitational field corresponding to this metric which is given by the Christoffel symbols
$$\Gamma_{10}^0={a'\over 2},\quad \Gamma_{00}^1={a'\over 2}e^{a-b}$$
$$\Gamma_{11}^1={b'\over 2},\quad \Gamma_{22}^1=-{r^2c'+2r\over 2}e^{c-b}$$
$$\Gamma_{33}^1=-{r^2c'+2r\over 2}e^{c-b}$$
$$\Gamma_{12}^2={c'\over 2}+{1\over r},\quad \Gamma_{33}^2=-\sin\te\cos\te$$
$$\Gamma_{13}^3={c'\over 2}+{1\over r},\quad \Gamma_{23}^3=\cot\te.\eqno(2.3)$$
Here the prime means $d/dr$, all other Christoffels vanish.

To simplify the following discussion we assume that the astronomer on earth has corrected his measurements for the motion of the earth with respect to the center of the galaxy, so that we can choose the center of the galaxy as origin of the laboratory coordinate system. Now the star moves on a geodesic (1.2)
$${d^2x^\mu\over d\tau^2}+\Gamma^\mu_{\al\beta}{dx^\al\over d\tau}{d x^\beta\over d\tau}=0.\eqno(2.4)$$
We consider the motion in the equatorial plane $\te=\pi/2$. Then we have to solve the following three differential equations:
$${d^2t\over d\tau^2}+a'{dt\over d\tau}{dr\over d\tau}=0\eqno(2.5)$$
$${d^2r\over d\tau^2}+{a'\over 2}e^{a-b}\B({dt\over d\tau}\B)^2+{b'\over 2}\B({dr\over d\tau}\B)^2-re^{c-b}\B({r\over2}c'+1\B)\B({d\phi\over d\tau}\B)^2=0
\eqno(2.6)$$
$${d^2\phi\over d\tau^2}+\B({2\over r}+c'\B){dr\over d\tau}{d\phi\over d\tau}=0.\eqno(2.7)$$
We want to find integrating factors for these three equations. Indeed multiplying (2.5) by $\exp a$ we get
$${d\over d\tau}\B(e^a{dt\over d\tau}\B)=0$$
so that
$$e^a{dt\over d\tau}=\cost=A$$
and
$${dt\over d\tau}=Ae^{-a}.\eqno(2.8)$$

Equation (2.7) is multiplied by $r^2$ which gives
$${d\over d\tau}\B(r^2{d\phi\over d\tau}\B)+c'{dr\over d\tau}r^2{d\phi\over d\tau}=0.$$
Dividing this by $r^2 d\phi/d\tau$ leads to
$${d\over d\tau}\log\B(r^2{d\phi\over d\tau}\B)+{dc(r)\over d\tau}=0.$$
After integration we obtain
$$\log\B(r^2{d\phi\over d\tau}\B)=-c+\cost$$
so that finally
$${d\phi\over d\tau}={J\over r^2}e^{-c}.\eqno(2.9)$$
Here the integration constant is chosen in such a way that $J$ reminds of the angular momentum in the standard theory. Finally we substitute (2.8) and (2.9) into (2.6). The resulting equation can be written in the form
$${d^2r\over d\tau^2}+{b'\over 2}\B({dr\over d\tau}\B)^2+{A^2\over 2}a'e^{-a-b}-{J^2\over r^3}e^{-b-c}\B({r\over 2}c'+1\B)=0.\eqno(2.10)$$
Here multiplication by
$$2e^b{dr\over d\tau}$$
yields the integrable equation
$${d\over d\tau}\B[e^b\B({dr\over d\tau}\B)^2\B]+A^2a'{dr\over d\tau}e^{-a}-{J^2\over r^3}e^{-c}{dr\over d\tau}(rc'+2)=0.\eqno(2.11)$$
After integration we have
$$e^b\B({dr\over d\tau}\B)^2-A^2e^{-a}+{J^2\over r^2}e^{-c}=\cost=B.\eqno(2.12)$$

The 3-velocity appearing in (2.1) which is measured by the astronomers is equal to
$$\vec V=\B({dx^1\over dt},{dx^2\over dt},{dx^3\over dt}\B).\eqno(2.13)$$
Since we consider motion in the equatorial plane $\te=\pi/2$, only the first and third components are different from zero. To eliminate the affine parameter $\tau$ in favor of the measured time $t$ we multiply by appropriate powers of
$${d\tau\over dt}={e^a\over A}.\eqno(2.14)$$
Then from (2.12) we get
$$e^b\B({dr\over dt}\B)^2=e^a-{J^2\over A^2}{e^{2a-c}\over r^2}+{B\over A^2}e^{2a}.\eqno(2.15)$$
In the following we are interested in the square
$$\vec V^2=-g_{11}\B({dr\over dt}\B)^2-g_{33}\B({d\phi\over dt}\B)^2=$$
$$=e^b\B({dr\over dt}\B)^2+{J^2\over A^2r^2}e^{2a-c}.\eqno(2.16)$$
Inserting (2.15) the term with $J^2$ drops out and we end up with the simple result
$$\vec V^2=e^a+{B\over A^2}e^{2a}.\eqno(2.17)$$

The result (2.17) is not yet the desired rotation velocity because the integration constants $A$ and $B$ must still be determined. To do so
we specialize everything for circular motion $r=\cost$ For $dr/dt=0$ in (2.15) we get the equation
$${J^2\over A^2}{e^{-c}\over r^2}-e^{-a}-{B\over A^2}=0.\eqno(2.18)$$
A second equation is obtained by differentiating this equation with respect to $r$ which is the stability condition for the circular path:
$$-2{J^2\over A^2}{e^{-c}\over r^3}-{J^2\over A^2}{c'e^{-c}\over r^2}+a'e^{-a}=0.\eqno(2.19)$$
This gives the following values for the integration constants
$${J^2\over A^2}={a'r^3\over rc'+2}e^{c-a}\eqno(2.20)$$
$${B\over A^2}={ra'\over rc'+2}e^{-a}-e^{-a}.\eqno(2.21)$$
Here $r$ now stands for the constant radius of the circular orbit. Now we are able to compute the circular velocity squared from (2.17)
$$\vec V_c^2\equiv w={ra'\over rc'+2}e^a.\eqno(2.22)$$

For a check we specialize the result (2.22) for the Schwarzschild metric where we have
$$a=\log\B(1-{r_s\over r}\B),\quad c=0\eqno(2.23)$$
and $r_s$ is the Schwarzschild radius
$$r_s=2MG\eqno(2.24)$$
with $M$ being the central point mass and $G$ Newton's constant. Then  $\vec V_c^2$ becomes
$$V_c^2={r_s\over 2r}={MG\over r}.\eqno(2.25)$$
This exactly coincides with Newton's theory. This circular velocity is right on the scale of the solar system. But on the scale of galaxies it is obviously not, $V_c(r)$ becomes constant for large $r$ instead of decreasing like $r^{-1/2}$. If one keeps to the Schwarzschild metric one must postulate some dark matter everywhere in the outer part of the galaxies. Without dark matter not only the Schwarzschild solution but also Newton's theory breaks down on large scales.

It was our program to determine the metric from the observable (2.22). To carry this through we must now solve Einstein's equation.

\section{Solution of the vacuum equation}

 The metric functions $a, b, c$ appearing in (2.2) must satisfy differential equation which follow from Einstein's equations. In standard general relativity one puts $c=0$. This is a special choice of gauge which leads to Birkhoff's theorem and the Schwarzschild metric. This works well in the solar system, but obviously not on the galactic scale.
The standard way out is to abandon the vacuum equations and assume some hypothetical dark matter. As long as this dark matter is not convincingly recorded one should also study the other possibility of retaining $c(r)\ne 0$. Then the vacuum solution is no longer unique. To fix it uniquely we take the expression (2.22) for the circular velocity $V(r)$ as our nonstandard gauge condition. It is often argued that by a transformation of coordinates $c=0$ can always be achieved.
We show at the end of this section (3.20) that one loses the contact to physics in this way.

Since the circular velocity $V(r)$ must be given the theory seems to have less predictive power. What seems to be a weakness is a strength: The asymptotic $V(r)$ cannot be predicted on the basis of the vacuum equations alone, the dynamics of the normal matter, that means the detailed structure of the galaxy, must necessarily be taken into account. Indeed a universal asymptotic velocity profile for all galaxies seems not to exist. In addition, only with $c\ne 0$ is it possible to carry out our program to express the metric by the observable $V(r)$. We continue the discussion of the nonstandard gauge in the concluding remarks.

The non-vanishing components of the Ricci tensor for the metric (2.2) are the diagonal elements 
$$R_{tt}={1\over 2} e^{a-b}(a''+{1\over 2} a'^2-{1\over 2} a'b'+a'c'+{2\over r}a')\eqno(3.1)$$
$$R_{rr}=-{1\over 2}(a''+2c'')+{b'\over 4}(a'+2c'+{4\over r})-{a'^2\over 4}-{c'^2\over 2}
-{2\over r}c'\eqno(3.2)$$
$$R_{\te\te}=e^{c-b}[-1-{r^2\over 2}c''-r(2c'+{a'-b'\over2})-{r^2\over4}c'(a'-b'+2c')]+1
\eqno(3.3)$$
$$R_{\phi\phi}=\sin^2\te R_{\te\te},\eqno(3.4)$$
the prime always denotes $\d /\d r$. Then the Einstein's equations without matter can be reduced to the following three differential equations  
$$G_{tt}=e^{a-b}\Bigl[-c''-{3\over 4}c'^2+{1\over 2}b'c'+{1\over r}(b'-3c')\Bigl]+{1\over r^2}(e^{a-c}-e^{a-b})=0\eqno(3.5)$$
$$G_{rr}={1\over 2}a'c'+{1\over r}(a'+c')+{c'^2\over 4}+{1\over r^2}\Bigl(1-e^{b-c}\Bigl)=0\eqno(3.6)$$
$$G_{\te\te}={r^2\over 2}e^{c-b}\Bigl[a''+c''-{1\over r}(b'-a'-2c')+{1\over 2}(a'^2-a'b'+a'c'-b'c'+c'^2)\Bigl]=0.\eqno(3.7)$$
As usual $G_{\al\beta}$ is the Einstein tensor.

It is not hard to see that there are only two independent field equations. Indeed, using (3.6) $b$ can be expressed by $a$ and $c$. Eliminating $b$ in (3.5) and (3.7) there results one second order differential equation for $a$ and $c$:
$$c''={a''\over a'}\Bigl(c'+{2\over r}\Bigl)+{4\over r^2}+a'c'+{c'^2\over 2}+{2\over r}(a'+c').\eqno(3.8)$$
Introducing the new metric function
$$f(r)=c(r)+2\log {r\over r_c}\eqno(3.9)$$
where $r_c$ has been included for dimensional reasons, equation (3.8) assumes the simple form
$${f''\over f'}-{a''\over a'}=a'+{f'\over 2}.\eqno(3.10)$$
This can immediately by integrated
$$\log{f'\over a'}=a+{f\over 2}+{\rm const.}\eqno(3.11)$$

On the other hand the circular velocity squared (2.22) becomes
$$V^2(r)\equiv w={a'\over f'}e^a.\eqno(3.12)$$
It is this velocity squared $w(r)$ which appears in all equations. Using (3.12) in (3.11) we have
$$f=-2\log w\eqno(3.13)$$
and
$$c=-2\log{rw\over r_c}\eqno(3.14)$$
where (3.9) has been used. This gives us the metric function
$$e^c=-g_{\te\te}r^{-2}={r_c^2\over r^2w^2}.\eqno(3.15)$$

To get $g_{tt}$ we return to (3.11) which can be written as
$$K_aa'e^a=f'e^{-f/2}.\eqno(3.16)$$
Here $K_a$ is the integration constant in (3.11). From (2.22) we find
$$a'e^a=w\B(c'+{2\over r}\B)={d\over dr}e^a.\eqno(3.17)$$
Combining this with (3.14)
$$c'=-2{w'\over w}-{2\over r}$$
we arrive at
$${d\over dr}e^a=-2w'.$$
This gives
$$g_{tt}=e^a=-2w+K_a.\eqno(3.18)$$

Finally $g_{\te\te}$ or $\exp b$ follows from (3.7). Solving for $\exp b$ we have
$$e^b=a'e^c\B({r^2\over 2}c'+r\B)+c'e^c\B({r^2\over 4}c'+r\B)+e^c.\eqno(3.19)$$
Substituting (3.18) and (3.15) we find
$$e^b=r_c^2{w'^2\over w^3}\B({1\over w}-{1\over w-K_a/2}\B).\eqno(3.20)$$
We choose the integration constants $K_a=1$ and $r_c=r_s/2$ where $r_s$ is the Schwarzschild radius (6.3.24). Then we get
$$e^c={r_s^2\over 4r^2w^2}\eqno(3.21)$$
$$e^a=-2w+1\eqno(3.22)$$
$$e^b={r_s^2\over 4}{w'^2\over w^4(1-2w)}.\eqno(3.23)$$
This reduces to the Schwarzschild solution ((3.25) below) for $w=r_s/2r$ (2.25).

Now we discuss again the subtle point of coordinate transformations. In other books the line element (2.2) is transformed to the so-called standard form by redefining the radial coordinate as follows
$$\bar r =re^{c/2}={r_s\over 2w}\eqno(3.24)$$
according to (3.21). Then our metric (2.2) assumes the Schwarzschild form
$$ds^2=\B(1-{r_s\over\bar r}\B)d\bar t^2-\B(1-{r_s\over\bar r}\B)^{-1}d\bar r^2-\bar r^2(d\te^2+\sin^2\te d\phi^2).
\eqno(3.25)$$
Mathematically the class of nonstandard solutions (3.21-23) has collapsed to the Schwarzschild solution. What does this mean physically ? As was repeatedly emphasized we reject to interpret the metric physically. Instead we consider the observable $w(r)=V_c^2(r)$. From (3.24) we obtain
$$w(r)={r_s\over 2\bar r}\equiv\bar w(\bar r).\eqno(3.26)$$
This is just the Schwarzschild expression in the new coordinate $\bar r$. Now it is clear what has been done : The new radius $\bar r$ has been chosen in such a way that the measured $w(r)$ becomes equal to the Schwarzschild expression $\bar w(\bar r)$. Such a transformation is trivially possible, but it has no physical significance. We see that {\it solutions that are equivalent under diffeomorphisms can be physically in-equivalent.} The reason  is that the physical observables transform non-trivially under coordinate transformations. One may ask the question: What is the right physical radius, $r$ or $\bar r$ ? The astronomer must give the answer. If he would work with $\bar r$ then for every measured rotation curve, i.e. for every galaxy, he must define a new radial coordinate $\bar r$. This is not what he does. He always applies the same measuring procedure (for example measuring the apparent luminosity) to all galaxies, and this gives our radius $r$. After all in reality, the astronomer adds, the rotation curves in galaxies are not Schwarzschild.

As far as the vacuum equations are concerned we are not able to predict the circular velocity; it must be given. But then
from (3.21-23) we are able to predict other observable quantities which can be computed from the metric, for example lensing data [8]. In this way the theory can be tested. Another test is investigated in the next section.

\section{Thin material disk with a dark halo}

We study a simple model of a spiral galaxy by assuming that the normal matter is concentrated in the equatorial plane $z=0$ with a singular density $\sim\delta^1(z)$. For this problem the theory of distribution valued curvature tensor is appropriate which is mainly due to Israel [9]. To be self-contained we give a simple derivation of the relations we need. Another reason to do this is the following: {\it In nonstandard general relativity we do not use geometric relations involving the metric.} Einstein's equation is the only basis, therefore, all derivations must be double checked.
Let $S$ be a three-dimensional surface in 4-space where the metric tensor $g_{\mu\nu}$ is continuous but has finite jumps in the normal derivatives; the derivatives in the tangential directions are assumed to be continuous. In an admissible coordinate system let $S$ be described by the equation
$$\fii(x)=0\eqno(4.1)$$
and have the normal vector
$$n_\mu={\d\fii\over\d x^\mu}.\eqno(4.2)$$
Then the finite discontinuities in the first partial derivatives of $g_{\mu\nu}$ are given by
$$[g_{\mu\nu,\sigma}]\equiv{\d g_{\mu\nu}\over\d x^\sigma}\Bigl\vert_+-{\d g_{\mu\nu}\over\d x^\sigma}\Bigl\vert_-
=n_\sigma b_{\mu\nu},\eqno(4.3)$$
where $+$ and $-$ mean the limiting values from both sides of $S$. This follows from the decomposition of the gradient into normal and tangential components. The corresponding jumps in the Christoffel symbols then are
$$2[\Gamma^\al_{\beta\gamma}]=n_\beta b^\al_\gamma+n_\gamma b^\al_\beta-n^\al b_{\beta\gamma}.\eqno(4.4)$$

The Ricci tensor
$$R_{\mu\nu}=\d_\al\Gamma^\al_{\mu\nu}-\d_\nu\Gamma^\al_{\mu\al}+\Gamma^\al_{\al\beta}\Gamma^\beta_{\mu\nu}-
\Gamma^\al_{\nu\beta}\Gamma^\beta_{\al\mu}\eqno(4.5)$$
contains derivatives of $\Gamma$, consequently the finite jumps lead to singular contributions proportional to the delta distribution $\delta_S$ with support on $S$ according to 
$$\d_\beta\Gamma^\al_{\mu\nu}\vert_{\rm sing}=[\Gamma^\al_{\mu\nu}]n_\beta\delta_S.\eqno(4.6)$$
Then it follows from (4.4) that
$$R_{\mu\nu}\vert_{\rm sing}={1\over 2}(-n^\al n_\al b_{\mu\nu}+n^\al\tilde b_{\mu\al}n_\nu+n^\al\tilde b_{\al\nu}n_\mu)\delta_S,\eqno(4.7)$$
with
$$\tilde b^\al_\beta=b^\al_\beta-{1\over 2}b\delta^\al_\beta,\quad b=g^{\mu\nu}b_{\mu\nu}.\eqno(4.8)$$
This is in agreement with eq.(2.14) of Taub [9], note that his convention for the Ricci tensor is the negative of our (4.5). 

In the Einstein's equations these singular distribution must be compensated by a distribution valued energy-momentum tensor
$$(R_{\mu\nu}-{1\over 2}R)g_{\mu\nu}\vert_{\rm sing}=\kappa t_{\mu\nu}\delta_S,\eqno(4.9)$$
where
$$R=g^{\al\beta}R_{\al\beta},\quad \kappa={8\pi G\over c^2}.\eqno(4.10)$$
If the jumps $b_{\mu\nu}$ of the normal derivatives of $g_{\mu\nu}$ are known, $t_{\mu\nu}$ can be calculated from
(4.7) and (4.9):
$$-2\kappa t_{\mu\nu}=n^2\Bigl((g^\sigma_\mu-{n^\sigma n_\mu\over n^2})(g^\tau_\nu-{n^\tau n_\nu\over n^2})-$$
$$-(g_{\mu\nu}-{n_\mu n_\nu\over n^2})(g^{\sigma\tau}-{n^\sigma n^\tau\over n^2})\Bigl)b_{\sigma\tau},\eqno(4.11)$$
where $n^2=n^\al n_\al$. This agrees with eq.(6-2) of Taub. The singular contribution (4.11) must be added to the regular energy-momentum tensor which renders the field equations fulfilled outside of the surface $S$.

Now we come to our simple galaxy model where the normal matter is concentrated in the plane $\te=\pi/2$ which is our singular surface $S$. Outside this plane we have vacuum with a dark halo as it is described by the nonstandard spherically symmetric solution (3.16-18). To have a simple representation of the plane $z=0$ and the corresponding delta-measure we go over to cylindrical coordinates $(t,R,z,\phi)$
$$r^2=R^2+z^2,\quad z=r\cos\te,\quad \sin\te={R\over r}.\eqno(4.12)$$
Then the metric (2.12) assumes the following non-diagonal form
$$ds^2=g_{\mu\nu}dx^\mu dx^\nu$$
with
$$g_{00}=e^a,\quad g_{11}=-{R^2\over r^2}e^b-{z^2\over r^2}e^c,\quad g_{22}=-{R^2\over r^2}e^c-{z^2\over r^2}e^b$$
$$g_{12}=g_{21}=-2{rz\over r^2}(e^b-e^c),\quad g_{33}=-R^2e^c.\eqno(4.13)$$
For simplicity we still write $r$, but our admissible coordinates are $x^1=R, x^2=z$. We also need the inverse
$$g^{00}=e^{-a},\quad g^{11}={g_{22}\over D},\quad g^{22}={g_{11}\over D}$$
$$g^{12}=-{g_{12}\over D}=g^{21},\quad g^{33}={1\over g_{33}},\eqno(4.14)$$
where the determinant $D$ is equal to
$$D=g_{11} g_{22}-(g_{12})^2=e^{b+c}-3{R^2z^2\over r^4}(e^b-e^c)^2.\eqno(4.15)$$

To construct the metric with the material disk we apply the widely used displace, cut, and reflect method which goes back to Kuzmin [10] and since then has been used by many authors. Following the procedure of Voigt and Letelier [11] we take the metric (3.7) in the half space $z>d>0$, displace it to $z=0$ and reflect it for $z<0$. This produces the finite jumps in the $z$-derivatives of $g_{\mu\nu}$. The whole procedure is equivalent to the transformation $z\to |z|+d$.  The normal vector is $n_\mu=(0,0,1,0)=\delta^2_\mu$ and
$$n^\nu=g^{\nu\mu}n_\mu=g^{\nu 2},\quad n^\nu n_\nu=g^{22}.$$
The jumps (4.3) in the normal derivatives on $z=0$ which we need are equal to
$$b_{11}=[g_{11, 2}]=g'_{11}{2d\over r}-{4d\over r^2}e^c\eqno(4.16)$$
$$b_{33}=[g_{33, 2}]=g'_{33}{2d\over r},$$
where the prime always means $\d /\d r$ keeping $z$ and $R$ constant. Now from (4.11) we find the energy density
$$t_0^0={1\over 2\kappa}\Bigl(Db_{11}+{g_{11}\over Dg_{33}}b_{33}\Bigl)\eqno(4.17)$$
with $D=g_{11}g_{22}-g_{12}^2$. Using
$$b_{11}={2d\over r}\Bigl(g'_{11}-{2\over r}e^c\Bigl),\quad b_{33}={2d\over r}g_{33}c',$$
we finally obtain
$$t_0^0=-{d\over\kappa r}\Bigl(e^{b+c}{R^6\over r^4}\d_r({e^b\over r^2})+{2R^4\over r^5}e^{b+2c}-{r^2\over R^2}\d_re^{-c}\Bigl).\eqno(4.18)$$
Here we have to put $z=0$ everywhere which gives $r^2=R^2+d^2$. 

 Now we must specify the circular velocity squared $u(r)$ in order to fix the metric. We are particularly interested in the case of an asymptotically flat circular velocity which in the usual terminology corresponds to a dark halo. Therefore we assume $u(r)$ of the form
$$u(r)=u_{\rm flat}+{u_1\over r}+O(r^{-2})\eqno(4.19)$$
for large $r$.  Then it follows from (3.16-18)
$$e^a=K_a+O(r^{-1}),\quad e^b={L_b\over r^4}+O(r^{-5})$$
$$e^c={L_c\over r^2}+O(r^{-3})\eqno(4.20)$$
 where by (3.17)
$$L_c\sim u_{\rm flat}^{-2}=V_{\rm flat}^{-4}.\eqno(4.21)$$
 Using this in (4.18) the leading order comes from the last term
$$t_0^0={2d\over\kappa L_c}{r^2\over R^2}(1+O(R^{-1})).\eqno(4.22)$$
This is proportional to the density of normal matter because we consider a static energy-momentum tensor. Taking (4.21) into account we find that
$$t_0^0\sim u^2_{\rm flat}\sim V^4_{\rm flat}(R)\eqno(4.23)$$
for large $R$. This is in accordance with the baryonic Tully-Fisher relation for galaxies [3] [4], which states that the total baryonic mass $M$ is proportional to $V^4_{\rm flat}$. In fact, the contribution of the inner part $R<R_1$ of the disk can be made arbitrarily small compared to the outer part between $R_1<R<R_2$, say [12]. We emphasize that $M$ is obtained from $t_0^0$ by integrating with the Euclidean surface measure $R\, dR\, d\phi$, because this is what astronomers are doing when they determine $M$ from luminosity measurements. Our theory gives a very natural explanation of the Tully-Fisher relation which , otherwise, theoretically and observationally is somewhat mysterious.
  
The radial pressure $t^r_r$ vanishes because $G_{rr}$ (3.6) does not contain a second derivative. Therefore our model must be interpreted as a dust disk with purely azimuthal stresses. This is not very realistic and it remains to be investigated whether the Tully-Fisher relation is a generic property for more physical galaxy models.

\section{Concluding remarks}

Our finding is that in the solar system the right gauge is $c(r)=0$, but on the galactic scale we have $c(r)\ne 0$. One would like to have a deeper understanding of this apparent paradox. One possible explanation is the following. At the very end general relativity must describe the solar system, the milky way, the local galaxy cluster etc. {\it simultaneously}. The division into separated subsystems is a misleading simplification. Keeping this in mind a continuous transition from $c$ approximately zero on small scales to $c\ne 0$ on the large is quite natural.

Obviously on small scales as the solar system or the binary pulsars the standard theory based on the geometric interpretation is the right one. But on the galactic scale which is a factor $10^8$ bigger the non-geometric aspect of general relativity becomes visible. In both cases we are observing geodesics in a gravitational field. On the small scale this field can be described geometrically, on the large scale this is not the appropriate picture.

Our solutions in Sect.3 seem to be the right ones to describe the dark halo of galaxies, if some dark matter cannot be found experimentally. The Tully-Fisher relation found in the last section is a central relation in modified Newtonian dynamics (MOND) [13]. This suggests that nonstandard GR is in accordance with MOND in contrast to standard GR. As far as the vacuum equations are concerned this is obviously true because nonstandard GR does not predict the circular velocity $V(r)$. The same remains true if we include normal matter in hydrostatic equilibrium [14]. We expect that the analysis of a detailed galaxy model in the framework of nonstandard GR will give the rotation curve $V(r)$. Indeed, the analysis of the last section shows that nonstandard GR solves the inverse problem: Given the rotation curve we can calculate the energy-momentum tensor. In standard GR the problem usually is posed the other way around. In the literature one has studied various modifications of GR to make MOND relativistic [15]. We have seen that this is not needed, nonstandard GR does the job.

\end{document}